\documentclass[12pt,a4paper]{article}
\usepackage{epsfig}
\begin{document}
\title{Detecting the non-universal gauge boson $Z^{'}$ at high-energy $e^{+}e^{-}$ collider}
\author{Chongxing Yue and Dongqi Yu \\
%\address{
{\small  Department of Physics, Liaoning Normal University, Dalian
116029, China}\thanks{E-mail:cxyue@lnnu.edu.cn}\\}
\date{\today}
      \maketitle

\begin{abstract}
\hspace{5mm}In the context of topcolor-assisted techicolor($TC2$)
models, we investigate the feasibility of detecting the
non-universal gauge boson $Z^{'}$ in the future high-energy linear
$e^{+}e^{-}$ collider($LC$) experiments by performing $\chi^{2}$
analysis via studying its virtual effects on the processes
$e^{+}e^{-}\rightarrow f\overline{f}$ with $f=\tau$, $\mu$, b, and
c. We find that the non-universal gauge boson $Z^{'}$ is most
sensitive to the process $e^{+}e^{-}\rightarrow
\tau\overline{\tau}$. Discovery limits of the $Z^{'}$ mass
$M_{Z^{'}}$ can be enhanced by the suitably polarized beams. If we
assume that polarization of electron beam and positron beam are
$80\%$ and $60\%$, the $LC$ experiment with $\sqrt{s}=500GeV$ and
$\pounds_{int}=340fb^{-1}$ can explore $M_{Z^{'}}$  up to 8$TeV$
for the coupling parameter $K_{1}\leq$0.8 at $95\%$ $CL$.
\end {abstract}

\vspace{2.0cm} \noindent
 {\bf PACS number(s)}:12.60Fr, 14.80.Mz, 12.15.Lk

\newpage

Although the standard model($SM$) has been successful in
describing the physics of electroweak interaction, it is only
considered as the low energy limit of a more fundamental theory
which is characterized by a large energy scale $\Lambda$.  New
Physics(NP) should exist at energy scales around $TeV$. Many
models of NP beyond the $SM$, such as string theory[1], grand
unified theory[2], strong top dynamical models[3], predict the
presence of massive $Z^{'}$ gauge bosons. If these new particles
are discovered, they would represent irrefutable proof of NP, most
likely that the $SM$ gauge groups must be extended[4]. Thus,
search for extra neutral gauge bosons $Z^{'}$ provides a common
tool in quest for NP at high energy colliders.

The hadron colliders, such as Tevatron and future $LHC$, are
expected to directly probe possible NP beyond the $SM$ up to a
scale of a few $TeV$, while a high-energy linear $e^{+}e^{-}$
collider($LC$) is required to complement the probe of the new
particles with detailed measurement. Even if their masses exceed
the center-of-mass(c.m.) energy $\sqrt{s}$, the $LC$ experiments
also retain all indirect sensitivity through a precision study of
their virtual corrections to observables. Thus, a $LC$ has a large
potential for discovery of new particles.

Now the interesting possibility is to investigate the feasibility
of a $Z^{'}$ discovery at the future $LC$ experiments. The
possible signals of extra gauge bosons $Z^{'}$ at $LC$ can arise
from the indirect effects of $Z^{'}$ exchange via the process
$e^{+}e^{-}\rightarrow f\overline{f}$. Through their interference
with the $SM$ $\gamma$ exchange and $Z$ exchange, significant
deviations from the $SM$ predictions can occur even when the
$Z^{'}$ mass $M_{Z^{'}}$ is much larger than the c.m. energy
$\sqrt{s}$[5]. The $LC$ observables, such as the cross section for
$f\overline{f}$ final state, forward-backward asymmetry
$A^{f}_{FB}$, and the left-right asymmetry $A^{f}_{LR}$, can give
information on the $Z^{'}$ parameters. Even if there is no
observed signal within the experimental accuracy, limits on the
free parameters of extra gauge boson $Z^{'}$ at a conventionally
defined confidence level($CL$) can be derived.

Ref.[6] has studied the virtual effects of extra gauge bosons
$Z^{'}$ on the process $e^{+}e^{-}\rightarrow f\overline{f}$ in
the context of the $\chi$ model occurring in the breaking
$SO(10)\rightarrow SU(5)\times U(1)_{\chi}$, the $\psi$ model
originating in $E_{6}\rightarrow SO(10)\times U(1)_{\psi}$, the
$\eta$ model, and the left-right model. The search limits for
these new neutral gauge bosons $Z^{'}$ at the present and future
$LC$ experiments with initial beam polarization are given by
studying some observables deviations from the $SM$ predictions. In
this paper, we discuss the possibility of detecting extra neutral
gauge bosons $Z^{'}$, predicted by topcolor-assisted
technicolor($TC2$) models[7] and flavor-universal $TC2$ models[8],
via the processes $e^{+}e^{-}\rightarrow f\overline{f}$ with
$f=\tau$, $\mu$, b, and c at the future $LC$ experiments with
$\sqrt{s}=500GeV$ and both beams polarized. We find that the value
of the $Z^{'}$ mass $M_{Z^{'}}$ for a $Z^{'}$ discovery at the
future $LC$ experiments is most sensitive to the process
$e^{+}e^{-}\rightarrow \tau\overline{\tau}$. In most of the
parameter space, the $M_{Z^{'}}$ can be explored up to 6$TeV$ via
this process at the $LC$ experiments with the c.m. energy
$\sqrt{s}=500GeV$ and the integrated luminosity
$\pounds_{int}=340fb^{-1}$.

To completely avoid the problems, such as triviality and
unnaturaless arising from the elementary Higgs in the $SM$,
various kinds of dynamical models are proposed and among which
$TC2$ models and flavor-universal $TC2$ models are very
interesting because they can explain the large top quark mass and
provide possible dynamics of elecroweak symmetry
breaking(EWSB)[3]. A common feature of these models is that the
$SM$ gauge groups are extended at energy well above the weak
scale. Breaking of the extended groups to their diagonal
sub-groups produces non-universal massive gauge bosons $Z^{'}$[9].
These new particles treat the third generation fermions
differently from those in the first and second generations and
couple primarily to the third generation fermions.

The flavor-diagonal couplings of the non-universal gauge boson
$Z^{'}$ to fermions, which are related our calculation, can be
written as[3,7]:
\begin{eqnarray}
\hspace{-5cm}\pounds_{Z^{'}}&=&g_{1}cot\theta^{'}Z^{'}\cdot
(J^{1}_{Z^{'}}+J^{2}_{Z^{'}}+J^{3}_{Z^{'}})\nonumber\\
&=&g_{1}cot\theta^{'}Z^{'}_{\mu}\cdot(\frac{1}{6}\overline{b}_{L}
\gamma^{\mu}b_{L}-\frac{1}{3}\overline{b}_{R}\gamma^{\mu}
{b}_{R}-\frac{1}{2}\overline{\tau}_{L}\gamma^{\mu}
\tau_{L}-\overline{\tau}_{R}\gamma^{\mu}\tau_{R})-g_{1}tan
\theta^{'}Z^{'}_{\mu}\cdot\nonumber\\&&(\frac{1}{6}
\overline{c}_{L}\gamma^{\mu}c_{L}+\frac{2}{3}\overline{c}_{R}
\gamma^{\mu}c_{R}-\frac{1}{2}\overline{\mu}_{L}\gamma^{\mu}\mu_{L}
-\overline{\mu}_{R}\gamma^{\mu}\mu_{R}-\frac{1}{2}\overline{e}_{L}
\gamma^{\mu}e_{L}-\overline{e}_{R}\gamma^{\mu}e_{R}).
\end{eqnarray}
Where $g_{1}$ is the ordinary hypercharge gauge coupling constant,
$\theta^{'}$ is the mixing angle with
$tan\theta^{'}=\frac{g_{1}}{\sqrt{4\pi K_{1}}}$. To obtain the top
quark condensation and not form a $b\overline{b}$ condensation,
there must be $tan\theta^{'}\ll1$[7,8]. The currents
$J^{1}_{Z^{'}}$, $J^{2}_{Z^{'}}$, and $J^{3}_{Z^{'}}$ involve the
first, second, and third generation fermions, respectively.

Integrating out the heavy gauge boson $Z^{'}$, the above couplings
give rise to effective low-energy four fermions interactions,
which can in general be written as:
\begin{equation}
\pounds_{eff,Z^{'}}=-\frac{2\pi
K_{1}}{M^{2}_{Z^{'}}}(J^{1}_{Z^{'}}+J^{2}_{Z^{'}}+J^{3}_{Z^{'}})^{2},
\end{equation}
which can generate virtual corrections to the process
$e^{+}e^{-}\rightarrow f\overline{f}$. $M_{Z^{'}}$ is the mass of
the non-universal gauge boson $Z^{'}$. If we take $f=\tau, \mu, b$
and c, then the differential cross sections for these processes
are given in Born approximation by the s-channel $\gamma$, $Z$,
and $Z^{'}$ exchange. Neglecting fermion mass $m_{f}$ with respect
to the c.m. energy $\sqrt{s}$, they have the form[6,10]:
\begin{equation}
\frac{d\sigma(f\overline{f})}{dcos\theta}=\frac{3}{8}[(1+cos\theta)^{2}
\sigma_{+}(f\overline{f})+(1-cos\theta)^{2}\sigma_{-}(f\overline{f})],
\end{equation}
where $\theta$ is the angle between the incoming electron and the
outgoing fermion in the c.m. frame. The
$\sigma_{+}(f\overline{f})$ and $\sigma_{-}(f\overline{f})$ can be
expressed in terms of the helicity cross sections
$\sigma_{\alpha\beta}$ with $\alpha,\beta=L, R$:
\begin{equation}
\sigma_{+}(f\overline{f})=\frac{D}{4}[(1-P_{eff})\sigma_{LL}(f\overline{f})+
(1+P_{eff})\sigma_{RR}(f\overline{f})],
\end{equation}
\begin{equation}
\sigma_{-}(f\overline{f})=\frac{D}{4}[(1-P_{eff})\sigma_{LR}(f\overline{f})+
(1+P_{eff})\sigma_{RL}(f\overline{f})]
\end{equation}
with
\begin{equation}
P_{eff}=\frac{P_{e}-P_{\overline{e}}}{1-P_{e}P_{\overline{e}}},\hspace{1.5cm}
D=1-P_{e}P_{\overline{e}}.
\end{equation}
Where $P_{e}$ and $P_{\overline{e}}$ are the degrees of
longitudinal electron and positron polarization.

The helicity cross sections $\sigma_{\alpha\beta}(f\overline{f})$
can be written as:
\begin{equation}
\sigma_{\alpha\beta}(f\overline{f})=\frac{N_{C}A}{e^{4}}\mid
M_{\alpha\beta}(f\overline{f})\mid^{2},
\end{equation}
where $N_{C}\simeq3(1+\frac{\alpha_{s}}{\pi})$ for quarks,
$N_{C}=1$ for leptons, and $A=\frac{4\pi\alpha_{e}^{2}}{3s}$. The
helicity amplitudes $M_{\alpha\beta}(f\overline{f})$ can be
written as :
\begin{equation}
M_{\alpha\beta}=Q_{e}Q_{f}+g_{\alpha}^{e}g_{\beta}^{f}\chi_{Z}+
g_{\alpha}^{'e}g_{\beta}^{'f}\chi_{Z^{'}},
\end{equation}
where $\chi_{i}=\frac{s}{s-M^{2}_{i}+iM_{i}\Gamma_{i}}$ represent
the gauge boson $Z$ and $Z^{'}$ propagators,
$g_{L}^{f}=\frac{e}{S_{W}C_{W}}(I^{f}_{3L}-Q_{f}S^{2}_{W})$ and
$g^{f}_{R}=-\frac{e}{S_{W}C_{W}}Q_{f}S^{2}_{W}$ are the left-hand
and right-hand coupling constants of the $SM$ gauge boson $Z$ to
fermions, $ I^{f}_{3L}$ is the third component of isospin, and
$Q_{f}$ is the fermion electricharge.
$S^{2}_{W}=sin^{2}\theta_{W}$, $\theta_{W}$ is the Weinberg angle.
The coupling constants $g^{'e}_{\alpha}$ and $g^{'f}_{\beta}$ can
be easily extracted from Eq.(1) for the non-universal gauge boson
$Z^{'}$ predicted by $TC2$ models or flavor-universal $TC2$
models.

The cross sections $\sigma^{exp}(f\overline{f})$, which can be
directly detected at the $LC$ experiments, can be written as:
\begin{equation}
\sigma^{exp}(f\overline{f})=\frac{1}{4}[\sigma_{LL}(f\overline{f})+
\sigma_{LR}(f\overline{f})+\sigma_{RR}(f\overline{f})+\sigma_{RL}
(f\overline{f})]=\frac{N_{tot}^{exp}}{D\pounds_{int}\varepsilon}.
\end{equation}
$N^{exp}_{tot}=N_{L,F}+N_{R,F}+N_{L,B}+N_{R,B}$ is the total
number of events observed at the $LC$ experiments with polarized
beams, which has been defined in Ref.[6,10]. The parameter
$\varepsilon$ is the experimental efficiency for detecting the
final state fermions. In the following calculation, we will take
the commonly used reference values of the identification
efficiencies: $\varepsilon=95\%$ for $l\overline{l}$;
$\varepsilon=60\%$ for $b\overline{b}$; $\varepsilon=35\%$ for
$c\overline{c}$.

The virtual effects of extra gauge bosons $Z^{'}$ on the process
$e^{+}e^{-}\rightarrow f\overline{f}$ can be detected via
considering deviations of the measured observables from the $SM$
predictions induced by these new particles. The experimental
constraints on the $Z^{'}$ parameters, such as the mass and the
coupling constants, can be estimated by performing $\chi^{2}$
analysis, i.e. by comparing the mentioned deviations with the
expected experimental uncertainty including the statistical and
the systematic one.

For the cross sections $\sigma(f\overline{f})$, the $\chi^{2}$
function is defined as:
\begin{equation}
\chi^{2}=[\frac{\Delta\sigma(f\overline{f})}{\delta\sigma
(f\overline{f})}]^{2},
\end{equation}
where $\Delta\sigma(f\overline{f})=\sigma^{TC2}(f\overline{f})-
\sigma^{SM}(f\overline{f})$ and $\delta\sigma(f\overline{f})$ is
the expected experimental uncertainty about the cross section
$\sigma(f\overline{f})$ both including the statistical and
systematic uncertainties. The allowed values of the $Z^{'}$
parameters by observation of the deviation
$\Delta\sigma(f\overline{f})$ out of the expected experimental
uncertainty $\delta\sigma(f\overline{f})$ can be estimated by
imposing $\chi^{2}>\chi^{2}_{CL}$, where the actual value of
$\chi^{2}_{CL}$ specifies the desired $'$confidence$'$ level. In
this paper, we take the value $\chi^{2}_{CL}=3.84$ for $95\%CL$
for a one-parameter fit and give the observability bound on the
$Z^{'}$ mass $M_{Z^{'}}$ for the fixed value of the coupling
parameter $K_{1}$.

From Eq.(8), one can see that the contributions of the gauge boson
$Z^{'}$ to the amplitudes $M_{\alpha\beta}$ can be represented by
the combination of the product of couplings
$g^{'e}_{\alpha}$$g^{'f}_{\beta}$ with the propagator
$\chi_{Z^{'}}$. If we assume $\sqrt{s}=500GeV$, there is
$\sqrt{s}\ll M_{Z^{'}}$. In this case, only the interference of
the $SM$ term with the $Z^{'}$ exchange is important and
$\Delta\sigma(ff)$ with $f=\tau,$ $\mu$, $b$, and $c$ contributed
by $Z'$ exchange can be written as:
\begin{eqnarray}
\Delta\sigma(\tau\overline{\tau})&=&\frac{A}{2C^{2}_{W}}Re\{[9+
\frac{4\chi_{Z}}{C^{2}_{W}}[S^{2}_{W}+(-\frac{1}{2}+S^{2}_{W})
+\frac{1}{4S^{2}_{W}}(-\frac{1}{2}+S^{2}_{W})^{2}]]\chi_{Z^{'}}\},\\
\Delta\sigma(\mu\overline{\mu})&=&\frac{A\alpha_{e}}{2K_{1}C^{4}_{W}}
Re\{[9+\frac{4\chi_{Z}}{C^{2}_{W}}[S^{2}_{W}+(-\frac{1}{2}+S^{2}_{W})
+\frac{1}{4S^{2}_{W}}(-\frac{1}{2}+S^{2}_{W})^{2}]]\chi_{Z^{'}}\},\\
\Delta\sigma(b\overline{b})&=&\frac{N_{C}A}{6C^{2}_{W}}
Re\{[-3+\frac{\chi_{Z}}{C^{2}_{W}}[\frac{4S^{2}_{W}}{3}+\frac{2}{3}
(-\frac{1}{2}+S^{2}_{W})+2(-\frac{1}{2}+\frac{1}{3}S^{2}_{W})
\nonumber\\&&+\frac{1}{S^{2}_{W}}(-\frac{1}{2}+S^{2}_{W})(-\frac{1}{2}+
\frac{1}{3}S^{2}_{W})]]\chi_{Z^{'}}\},\\
\Delta\sigma(c\overline{c})&=&\frac{N_{C}A\alpha_{e}}{6K_{1}C^{4}_{W}}
Re\{[10+\frac{\chi_{Z}}{C^{2}_{W}}[-\frac{16S^{2}_{W}}{3}+2(\frac{1}{2}-
\frac{2}{3}S^{2}_{W})-\frac{8}{3}(-\frac{1}{2}+S^{2}_{W})\nonumber\\&&
+\frac{1}{S^{2}_{W}}(-\frac{1}{2}+S^{2}_{W})(\frac{1}{2}-\frac{2}{3}
S^{2}_{W})]]\chi_{Z^{'}}\}.
\end{eqnarray}

The square of the expected uncertainty about the cross section
$\sigma(f\overline{f})$ can be written as[10]:
\begin{equation}
[\delta\sigma(f\overline{f})]^{2}=\frac{[\sigma^{exp}(f\overline{f})
]^{2}}{N_{tot}^{exp}}+[\sigma^{exp}(f\overline{f})]^{2}
[\frac{P^{2}_{e}P^{2}_{\overline{e}}}{D^{2}}(\varepsilon^{2}_{e}
+\varepsilon^{2}_{\overline{e}})+\varepsilon^{2}_{\pounds}].
\end{equation}
The first term of Eq.(15) arises from the statistical uncertainty
and the second term represents the systematic uncertainty. In our
numerical estimation, we will take
$\varepsilon_{e}=\varepsilon_{\overline{e}}=\frac{\delta
P_{e}}{P_{e}}=\frac{\delta
P_{\overline{e}}}{P_{\overline{e}}}=0.5\%$,
$\varepsilon_{\pounds}=\frac{\delta\pounds_{int}}{\pounds_{int}}=0.5\%$.

It has been shown that, if the Landau pole of strong $U(1)$
interaction is to lie at least an order of magnitude above the
symmetry breaking scale $\Lambda$, the coupling parameter $K_{1}$
should satisfy certain constraint, i.e., $K_{1}\leq1$[8]. The
lower limits on the $Z^{'}$ mass $M_{Z^{'}}$ can be obtained via
studying its effects on various observables, which has been
precisely measured at the present collider experiments[3]. For
example, Ref.[9] has shown that to fit the electroweak measurement
data, the $Z^{'}$ mass $M_{Z^{'}}$ must be larger than 1$TeV$. The
lower bounds on $M_{Z^{'}}$ can also be obtained from dijet and
dilepton production at the Tevatron experiments[11], or from
$B\overline{B}$ mixing[12]. However, these bounds are significant
weaker than those from precision electroweak data. Ref.[8] has
simply considered the possible of detecting $Z'$ in the future
$LC$ experiments. They find that, for $M_{Z'}\leq2.7TeV$,
$K_{1}\leq1$, the signals of $Z'$ can be detected in the $LC$
experiments with $\sqrt{s}=500GeV$ and $\pounds=50fb^{-1}$. In
this paper, we Use above equations to investigate the upper limits
on the $Z^{'}$ parameters in the case of a $Z^{'}$ discovery in
the future $LC$ experiments with $\sqrt{s}=500GeV$ and obtain
$\pounds_{int}=340fb^{-1}$[5] and the range of the $Z'$ mass
$M_{Z'}$ which would be visible in the future $LC$ experiments for
the fixed value of the coupling parameter $K_{1}$.

\begin{figure}[htb]
\vspace{-0.5cm}
\begin{center}
\epsfig{file=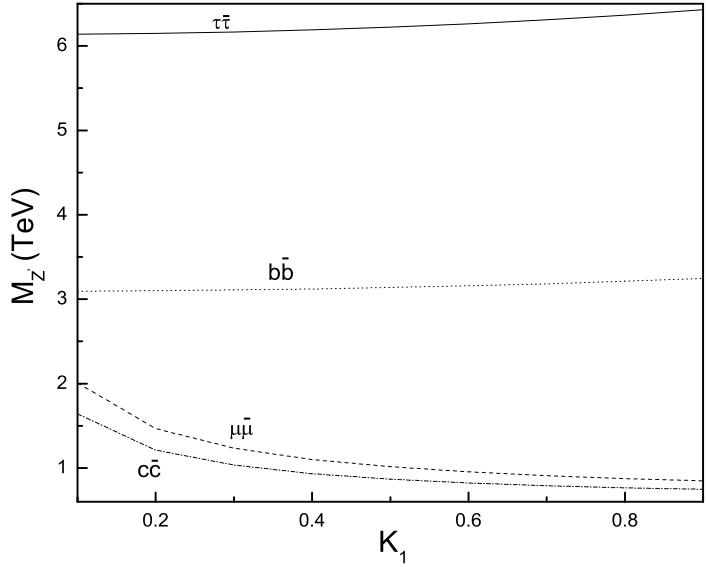,width=300pt,height=250pt} \vspace{-1.0cm}
\hspace{5mm} \caption{Searching limits of the $Z^{'}$ mass
$M_{Z^{'}}$ at 95$\%CL$ as function of the parameter
\hspace*{2.0cm}$K_{1}$ for
$\pounds_{int}=340fb^{-1}$,$(P_{e},P_{\overline{e}})=(0.8,0.6)$
and the processes $e^{+}e^{-}\rightarrow l\overline{l}$ with
\hspace*{2.0cm}$l=\tau,$ $\mu$, $b$, and c.} \label{ee}
\end{center}
\end{figure}

In our numerical estimation, we take the $SM$ parameters
$\alpha_{e}=\frac{1}{128.8}$, $M_{Z}=91.187GeV$,
$S^{2}_{W}=0.2315$, and $\Gamma_{Z}=2.495GeV$[13]. The total decay
width $\Gamma_{Z^{'}}$ of the gauge boson $Z^{'}$ is dominated by
$t\overline{t}$ and $b\overline{b}$, which can be written as:
$\Gamma_{Z^{'}}\approx\frac{K_{1}M_{Z^{'}}}{3}$[3].

Searching limits of the $Z^{'}$ mass $M_{Z^{'}}$ at $95\%CL$ are
plotted in Fig.1 as function of the coupling parameter $K_{1}$ for
the processes $e^{+}e^{-}\rightarrow l\overline{l}$ with $l=\tau,$
$\mu$, $b$, and $c$, in which we have assume $P_{e}=0.8$,
$P_{\overline{e}}=0.6$ and the integrated luminosity
$\pounds_{int}=340fb^{-1}$. From Fig.1, we can see that the ranges
of $M_{Z'}$ which would be visible in the $LC$ experiments are not
sensitive the parameter $K_{1}$. Discovery limits of $M_{Z'}$ for
the processes $e^{+}e^{-}\rightarrow\tau\overline{\tau}$ and
$e^{+}e^{-}\rightarrow b\overline{b}$ are larger than those
arising from the process $e^{+}e^{-}\rightarrow \mu\overline{\mu}$
or $e^{+}e^{-}\rightarrow c\overline{c}$. This is because the
gauge boson $Z^{'}$ couples primarily to the third generation
fermions. For $K_{1}=0.8$, the maximal values of the $Z^{'}$ mass
$M_{Z^{'}}$, which can be explored, are $6.3TeV$, $0.87TeV$,
$3.2TeV$, and $0.77TeV$ for the processes $e^{+}e^{-}\rightarrow
\tau\overline{\tau}$, $\mu\overline{\mu}$, $b\overline{b}$, and
$c\overline{c}$, respectively. Thus, the non-universal gauge boson
$Z^{'}$ is most sensitive to the process $e^{+}e^{-}\rightarrow
\tau\overline{\tau}$. The possible signals of $Z'$ might be more
easy detected via this process than via other processes in the
future $LC$ experiments. Certainly, if an new gauge boson $Z'$ is
indeed discovered in the future high energy experiments, we should
consider all of the possible processes to study in detail what
kind of $Z'$ one has found.

\begin{figure}[htb]
\vspace{-0.5cm}
\begin{center}
\epsfig{file=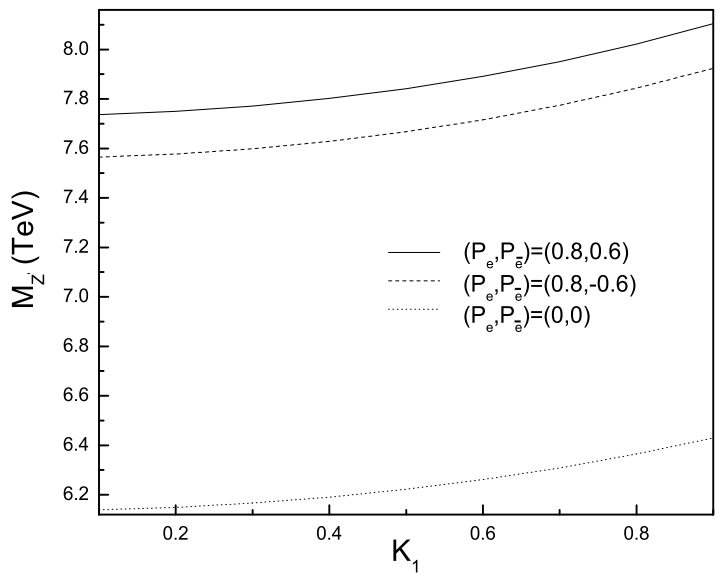,width=300pt,height=250pt} \vspace{-1.0cm}
\hspace{5mm} \caption{Searching limits of the $Z'$ mass $M_{Z'}$
at $95\%$CL as function of the coupling \hspace*{2.0cm}parameter
$K_{1}$ for $\pounds_{int}=340fb^{-1}$ and different beam
polarization.} \label{ee}
\end{center}
\end{figure}

A strong longitudinal polarization programme in the future $LC$
experiments with considerable polarization of the electron beam
and the positron beam is planned[14]. Beam polarization is not
only useful for a possible reduction of the background, but might
also serve as a possible tool to disentangle different
contributions to the signal. Beam polarization of the electron and
positron beams would lead to a substantial enhancement of the
cress section of some processes and thus enhance the $LC$
potential in the search for a new gauge boson $Z'$. To see the
effects of beam polarization on the searching limits of the $Z'$
mass $M_{Z'}$, we plot $M_{Z'}$ as a function of the coupling
parameter $K_{1}$ in Fig.2 for the process
$e^{+}e^{-}\rightarrow\tau\overline{\tau}$ for
$\pounds_{int}=340fb^{-1}$ and different beam polarization, in
which the solid line, dashed line, and dotted line represent
($P_{e},P_{\overline{e}}$)=(0.8,0.6), (0.8,-0.6), and (0,0),
respectively. From Eq.(15), we can see that the value of
$[\delta\sigma(f\overline{f})]^{2}$ for
$(P_{e},P_{\overline{e}})=(0.8,0.6)$ is same as that of
$(P_{e},P_{\overline{e}})=(-0.8,-0.6)$. Thus, we have not plotted
the curve line for $(P_{e},P_{\overline{e}})=(-0.8,-0.6)$ in
Fig.2. One can see from Fig.2 that the searching limits of the
$Z'$ mass $M_{Z'}$ are indeed sensitive to the polarization of
electron and positron beam. For $K_{1}=0.5$, the maximal values of
$M_{Z'}$ which can be detected in the $LC$ experiments via the
process $e^{+}e^{-}\rightarrow\tau\overline{\tau}$ are $7.84TeV$,
$7.67TeV$, and $6.22TeV$ for $(P_{e},P_{\overline{e}})=(0.8,0.6)$,
$(0.8,-0.6)$, and $(0,0)$, respectively.

From above formulas, we can see that discovery limits of the $Z'$
mass $M_{Z'}$ vary as the integrated luminosity $\pounds_{int}$
varying. For example, if we assume that the value of the
$\pounds_{int}$ increase from $100fb^{-1}$ to $350fb^{-1}$, the
discovery limits can increase from $7.15TeV$ to $7.67$ for
$K_{1}=0.5$, $(P_{e},P_{\overline{e}})=(0.8,0.6)$ and the process
$e^{+}e^{-}\rightarrow\tau\overline{\tau}$. Certainly, the change
in the LC reach with increasing luminosity is extremely small.

Strong top dynamical models, such as $TC2$ models and
flavor-universal $TC2$ models, predict the presence of the
non-universal gauge bosons $Z^{'}$, which couple preferentially to
the third generation fermions. These new particles can generate
significantly contributions to some observables. The lower bounds
on the mass or the coupling parameters of the gauge boson $Z^{'}$
have been derived from data taken at $LEP$ and Tevatron
experiments. In this paper, we investigate the feasibility of
$Z^{'}$ discovery in the future $LC$ experiments by performing
$\chi^{2}$ analysis. We focus our attention to the processes
$e^{+}e^{-}\rightarrow f\overline{f}$ with $f=\tau, \mu, b$ and c
in the context of $TC2$ models. We find that the non-universal
gauge boson $Z^{'}$ is most sensitive to the process
$e^{+}e^{-}\rightarrow \tau\overline{\tau}$. The $Z^{'}$ mass
$M_{Z^{'}}$ can be explored up to $7.7(7.8)TeV$ for the coupling
parameter $K_{1}=0.5(0.8)$ and ($P_{e},P_{\overline{e}}$)=(0.8,
-0.6) at the future $LC$ experiment with $\sqrt{s}=500GeV$ and
$\pounds=340fb^{-1}$. The observation upper limits of $M_{Z^{'}}$
can be further enhanced by the suitably polarized beams.

\vspace{1.5cm} \noindent{\bf Acknowledgments}

We thank the referee for carefully reading the manuscript. This
work was supported by the National Natural Science Foundation of
China (90203005) and the Natural Science Foundation of the
Liaoning Scientific Committee(20032101).

\end{document}